# A symmetric relaxation method for entire two-dimensional cellular networks and its implications

Kai Xu[a]*, Lifan Weng[b], Zihan Wang[b], Yuyang Lian[b], Bin Huang[b]
[a] *Fisheries College, Jimei University, Xiamen, 361021, China;*
[b] *College of Computer Engineering, Jimei University, Xiamen, 361021, China*
**Corresponding author: kaixu@jmu.edu.cn, kxu2013@gmail.com*

Kai Xu: http://orcid.org/0000-0002-1341-1525

**Abstract** To simulate the relaxation of an entire 2D cellular network, this study proposes a symmetric relaxation method for both inner and marginal vertices. The relaxations of these two types of vertices are determined by the central angle symmetry of associated cells and the angle symmetry at each vertex, but with different major considerations. Trimmed Voronoi networks with varying irregularity are used as initial networks for the relaxation simulation. In particular, we propose a regular hexagon disordering method to generate Voronoi networks and find that the inner cells of networks with an irregularity value of one exhibit a conserved edge number distribution, as found in other 2D cellular networks. Simulation results agree with the von Neumann-Mullins law for both inner and marginal cells, and a modified equation including a geometric correction term significantly improves prediction quality. The Aboav-Weaire law and Lewis law are also reproduced, with the latter showing that relaxed cells tend to approach the ellipses' maximum inscribed polygons. Analysis of edge length, interior angle, and shape index reveals that symmetric relaxation inhibits T1 (neighbour exchange) topological transitions by reducing short edges while increasing area disparity among neighbouring cells. The findings suggest that T1 events may be triggered when force disequilibrium overcomes the stabilising effect of symmetric relaxation, providing a possible mechanistic explanation for T1 in 2D foams.

## 1 Introduction

From the perspective of quantitative analysis, the dynamic behaviours of a 2D cellular network can be classified into two categories: the continuous geometric changes of individual cells and the discontinuous topological processes involving a few cells. The latter change the vertex number (which equals the edge number) of the related polygonal cells and thereby alter cell shape; the former only change vertex positions, while cell geometry remains constrained by topological parameters, such as the coordination number of a given vertex and the edge number of the cell. In real conditions, these two categories of behaviours intertwine and interact closely, making it difficult to distinguish and quantify their effects, summarise their patterns, and explore underlying mechanisms. The major geometric parameters used to quantify the shape of a cell include area, perimeter, edge length, interior angle, aspect ratio, and shape index. The shape index is defined as the ratio of perimeter to the square root of area; this parameter depends only on shape and remains unchanged under scaling and rotation (1). Apart from a vertex coordination number of three, two important topological parameters for an individual cell are its edge number $n$ and the average edge number $m$ of its neighbouring cells. These parameters comprehensively reflect cell shape, spatial position, as well as the sequence and direction of network evolution. Several laws have been summarized or derived through the investigation of these parameters, such as Lewis law, Aboav-Weaire law, and von Neumann-Mullins law (2, 3).

Under real conditions, typical 2D networks such as 2D foams and thalli of the seaweed *Pyropia haitanensis* usually have smooth boundaries and contain only a limited number of cells. The geometric and topological properties of marginal cells located on the network boundary differ significantly from those of inner cells that are fully surrounded by other cells (4-6). Marginal cells have an average edge number of approximately five, which is less than that of inner cells by one (6). In this study, vertices on the network boundary are defined as marginal vertices, and the rest are defined as inner vertices. For *Pyropia* tissue, the interior angles of inner vertices tend to be $120^{o}$, while those of marginal vertices tend to be 90° (6). Such distinct geometric features indicate that the dynamic behaviours of the two types of vertices are also different.

The vertex model has been widely used to simulate the evolution of a 2D cellular network. The simulation typically starts from a polygonal tessellation network, followed by vertex movement to relax the whole network. Topological transitions are triggered once edge length or cell area reaches preset thresholds (7-11). The topological processes break the equilibrium of the network and require further relaxation. In physics-based vertex models, the target positions of both marginal and inner vertices are calculated following the energy minimum principle, with some technical differences (12, 13). Recently, a geometry-based relaxation algorithm for inner vertices was established based on two geometric rules. This algorithm reproduced typical variation patterns of cell area and agreed with the von Neumann-Mullins law (14). In that study, the movement of each inner vertex was controlled by the central angle symmetry of regular polygons and was restricted by the angle symmetry at each vertex. The central angle symmetry sets a strong constraint on cell shape change because all vertices of an inner cell are coupled together, whereas the angle symmetry at each inner vertex drove interior angles toward $120^{o}$. However, the independent effects of the two geometric rules on cell shape remain unclear and need to be quantitatively analysed to reveal their mechanisms. In addition, given the significant difference between inner and marginal vertices regarding their associated interior angles, the above algorithm cannot be directly applied to marginal vertices.

To address the above research gaps, this study proposes a symmetric relaxation method for marginal vertices based on our observations on *Pyropia* tissue (marginal vertex-associated interior angles tend to be 90°) (6). Combined this approach with the existing relaxation algorithm for inner vertices (14) allow us to relax the entire network. We further compare and analyse the geometric behaviours of marginal and inner cells induced by relaxation, and discuss the possible mechanisms of the T1 (neighbour exchange) topological transition.

## 2 Methods

### *2.1 Initial networks*

Initial networks are constructed using Voronoi tessellations generated by randomly disturbing seed positions of regular hexagonal grids, with modifications based on a previous study (15). In this study, each seed is moved to a random position within its corresponding regular hexagon and this new method is named the regular hexagon disordering algorithm. The improved method avoids overlapping of the potential appearance zones of seeds by restricting each seed to its own regular hexagon. The coordinates of the disordered seed for the $i$th Voronoi cell are:

$$\begin{cases} X_i = X_{i0} + k \cdot dx \\ Y_i = Y_{i0} + k \cdot dy \end{cases}, \quad (1)$$

where ($X_{i0}$, $Y_{i0}$) are the initial coordinates of the $i$th seed; ($X_i$, $Y_i$) are the disturbed coordinates; $k$ denotes the irregularity of disordered Voronoi networks (15); $l$ is the edge length of regular hexagons; $dx$ and $dy$ are random disturbance components in the x (horizontal) and y (vertical) directions, respectively. The random point ($dx$, $dy$) is generated inside a regular hexagon of edge length $l$ (centered at the origin and with the same orientation as the grid hexagons) via a rhombus-based sampling method. Specifically, the unit regular hexagon (edge length = 1) is first divided into three rhombuses of equal area. One rhombus is randomly selected, and then a random point within it is calculated using two independent uniform random parameters $u, v \in [0,1]$. Finally, the point's coordinates are scaled by $l$ to match the target hexagon size. We then trim the disordered Voronoi network: if a marginal vertex is shared by only two edges, it is deleted and its two adjacent vertices are directly connected. This processing ensures that every vertex in the network is connected by three edges.

### *2.2 Symmetric relaxation method*

To fully equilibrate the entire network, the relaxation operation moves the inner vertices and marginal vertices to their target positions step by step. The target position of each inner vertex is calculated following our previous study (14). The calculation is based on $n$ optimal rays emanating from the centroid of an $n$-edged cell, where the angle between adjacent rays equals the central angle of a regular $n$-gon. The directions of these optimal rays are determined via least squares optimization to minimize the sum of squared angles between the rays and the lines connecting the centroid to each vertex. For each inner vertex, three corresponding optimal rays form a triangle, and the centroid of this triangle serves as the target position for relaxation (Figure 1A). For a marginal vertex, the relaxation scheme moves it along the marginal edge associated with its smaller marginal angle to the midpoint of this edge (Figure 1B).

Calculate the distance $D$ between the current position and the target position of every vertex. Before the start of an iteration, rank all distances in a descending order. In each iteration of relaxation, move all vertices one-by-one toward their target positions following the order. The motion distance of each movement is $\gamma$ times $D$. Before the movement of a vertex, update its target position and recalculate $D$, but still use the order. If a movement of a vertex results in any interior angles related to the vertex and neighbouring vertices becoming larger than 180°, do not execute the movement and skip to the next vertex. Additional constraints are applied to inner and marginal vertices. For an inner vertex, movement is skipped if it increases the sum of squares of its three associated angles or if the vertex is already positioned within the triangle (14). Without the former constraint, either the number of iterations must be substantially increased to fully relax the entire network, or the parameter $\gamma$ must be raised. The detailed differences arising from the inclusion or exclusion of this angle constraint are presented and discussed at Section 3. For a marginal vertex belonging to a cell with at least four edges, movement is skipped if the difference between its two marginal angles is less than 20°; otherwise (i.e., for a marginal vertex belonging to a three-edged cell), movement is skipped if the difference is less than 60°. The angle thresholds are set following the observations: about 80% of marginal angles in *Pyropia* marginal cells range from 80° to 100° (6) and triangles in 2D foams are very close to regular triangles.

In this study, an inner vertex's motion is mainly controlled by the symmetry of its three associated cells, and is also influenced by the angle constraint which represents the symmetry at the vertex where three cells meet. As for each marginal vertex, its motion is mainly controlled by the symmetry at the vertex where two marginal cells meet, and is also influenced by the symmetry of the two associated marginal cells. An inner cell contains only inner vertices, whereas each marginal cell contains, on average, two marginal vertices and three inner vertices. Consequently, the relaxation of marginal cells is quite different from that of inner cells. Considering that the movements of both inner and marginal vertices are determined by the central angle symmetry of associated cells and the angle symmetry at each vertex, the method is named symmetric relaxation.

***2.3 Simulation experiment and analysis***

To collect sufficient data, each trimmed Voronoi network was generated from 400 seeds at three $k$ values (0.2, 0.6, and 1) and at least 10 networks were generated at each $k$ value. To fully relax the entire Voronoi networks, we set $\gamma = 0.1$ and performed 100 iterations following our previous study (14). Fewer than 2% of vertices were moved at the 100th iteration; marginal vertices required far fewer iterations to achieve full relaxation compared with inner vertices. A comparison of the average area of 5- and 7-edged inner cells before and after relaxation revealed that 10 iterations achieved over 67% of the area change observed after 100 iterations. This yields a characteristic relaxation constant of 9.02 iterations, demonstrating that the network undergoes an exponential relaxation and ensuring that 100 iterations fully relax the network. For relaxation without the angle constraint, we set $\gamma = 0.2$ and performed 100 iterations, and fewer than 2% of vertices were moved at the 100th iteration. Figure 2 presents examples of trimmed Voronoi networks and their corresponding relaxed networks with and without the angle constraint.

The least squares method was used to fit an ellipse for each cell containing at least five vertices (14). For a cell with only three or four vertices, the LMG method in the R package *conicfit* (version 1.0.4) was applied (16). The minimum bounding rectangle of each cell was calculated using the Python package *pyenvelope*. Subsequently, the half-length and half-width of the rectangle, the inclination angle between its long side and the x-axis, together with the centroid coordinates of the cell, serve as initial input parameters for the ParGini function within the LMG method.

***2.4 2D foam observations***

Place a small volume (~4 $mm^3$) of shaving foam (Gillette Foamy) onto a glass microscope slide, then firmly flatten the foam by applying pressure with a coverslip. Let the sample evolve for about 5 minutes, then observe the evolution of the 2D foam under a microscope.

## 3 Results and discussion

***3.1 Cell edge number of trimmed Voronoi networks***

The topological and geometric features of the marginal cells located at the boundary of a 2D cellular network are quite different compared with those of inner cells (6). However, to date, there is no standard method to directly generate a bounded 2D network with the same mathematical characteristics as found in real networks. In this study, the Voronoi networks were trimmed to ensure that all vertices have a coordination number of three and were then used as the initial networks for the simulation. The present study used the regular hexagon disordering algorithm to generate Voronoi networks. An interesting finding is that, for the Voronoi network with an irregularity $k = 1$ (Figure 3A), its inner cells showed a very similar distribution of cell edge number compared with a non-living 2D network, the amorphous $SiO_2$ network (17) and two living 2D networks, the thallus of the seaweed *Pyropia* (6, 18) and the larval wing disc of the animal *Drosophila* (19-21). A previous study suggested that, for *Drosophila* epithelia, the specific distribution of edge number can be simulated by a Markov chain model of cell division regardless of initial distributions (19). However, this model should not apply to the 2D amorphous networks and random-disordered Voronoi networks. Because 2D Voronoi networks are mathematically defined, the conserved polygon distribution may stem from the mathematical constraints governing 2D space filling.

In addition, the network reverts to a regular hexagon grid as the irregularity decreases from 1 to 0. Similar phenomena have also been found in real conditions, for example, the percentage of hexagonal cells increased from ~45% to ~78% during the development of the *Drosophila* pupal wing (22). T1 topological transition has been considered as the primary driving mechanism for the increase in hexagonal cell percentage (22). In contrast, in this study, changes in seed positions contribute to the shift towards the regular hexagon grid.

For 2D networks with a smooth boundary, for example, *Pyropia* thalli and Bénard-Marangoni structures, their marginal cells are dominated by pentagons and the average edge number is approximately five (5, 6). For the trimmed Voronoi network in this study, the average edge number of marginal cells is approximately five regardless of the $k$ value, although the polygon distribution significantly changes with $k$ values (Figure 3B). Therefore, the trimmed Vonoroi networks are suitable for the simulation.

### *3.2 von Neumann-Mullins law*

For the $k$ values of 0.2, 0.6, and 1, the average area of initial inner cells was 0.86 and increased only slightly after relaxation (by less than 0.3%), while that of initial marginal cells was 0.68 and decreased by less than 1.7%. Our previous study found that the area variation of inner cells approximately follows the famous von Neumann-Mullins law (14). According to this law (23, 24), the area of an $n$-edged inner cell varies as

$$dA/_{dt} = C_{\ 1}(n-6), \quad (2)$$

where, $n$ is the edge number of the cell, 6 is the average edge number $\langle n \rangle$ of inner cells, $C_{\ 1}$ denotes a constant. Inner cells with more than six edges grow, cells with fewer than six edges shrink, and six-edged cells remain constant. It has been shown that the marginal cells located on the smooth boundary of a 2D cellular network have an average edge number of approximately five (6). Given the significant difference between inner and marginal cells on $\langle n \rangle$, the von Neumann-Mullins law should also be different. For a 2D network with a straight boundary, assuming angles between adjacent edges at inner and marginal vertices are 120° and 90°, respectively, Aref and Herdtle deduced the von Neumann-Mullins law under boundary conditions (25, 26)

$$dA/_{dt} = C_1(n-5). \quad (3)$$

This equation predicts that it is the pentagons that do not change their sizes. Then, for a 2D cellular network with a straight boundary, the von Neumann-Mullins law can be written as

$$dA/_{dt} = C_1(n-\langle n \rangle). \quad (4)$$

The simulation results for inner and marginal cells of this study and our previous study (14) agree with Eq. (4) except for $n = 3$ (Figure 4A, B). Here, we use the area difference $\Delta A$ between the Voronoi cells and their fully relaxed counterparts to represent the area growth rate.

The von Neumann-Mullins law was derived for 2D foam based on two assumptions: cell wall (edge) velocity is proportional to its mean curvature and the vertex angles of inner and marginal vertices are 120° and 90°, respectively. However, these two assumptions are not used in our model. Similar simulation results for inner cells have been reported based on a physics-based vertex model in which all the edges are also straight (11, 27). This vertex model includes energy-driven vertex motion together with both T1 and T2 (disappearance of a triangle) topological transitions, allowing full simulation of the evolution of 2D foams. However, our model does not contain T1 and T2 which is a significant difference. The straight edges of few-edged cells have been attributed to the deviation from the von Neumann-Mullins law (11). By introducing symmetric cell configurations into the vertex motion equation, Nakashima et al. (1989) analytically derived the von Neumann-Mullins equation for inner cells (11). On this basis, the symmetry-based method adopted in our earlier study (14) shares conceptual connections with the physics-based vertex motion equation used in earlier studies (11, 27). In particular, the results of this study and previous studies (11, 14, 27) suggest that the von Neumann-Mullins law does not rely on topological processes.

In this and our previous study (14), the inner vertices move following the rule of central angle symmetry, but may be disturbed by an angle constraint: the interior angle symmetry at vertices. This constraint drives the interior angles toward 120°, which differs from the rule governing vertex motion. This study then compared the effects of relaxation on cells with and without this angle constraint. For both inner and marginal cells, addition of the constraint showed little effect on the area variation for $n \geq 4$, while significantly improved the simulation results of 3-edged cells (Figure 4A-B).

In addition, for inner cells at $k = 1$, relaxation without the constraint strongly concentrated the edge length in the range of 0.5 to 0.7 (Figure 4C), but resulted in a significantly dispersed angle distribution (Figure 4D). Compared with the relaxation without the constraint, the addition of the constraint decreased the proportion of edge lengths within 0.5-0.7 from 89% to 63%, while increasing the proportion of angles within 110°-130° from 26% to 46%. The interior angles of *Pyropia* inner cells tend to shift toward 120° (18), while 2D foam cells have been observed to be very close to ellipses' maximal inscribed polygons (EMIP) (28). Similar results also can be found on marginal cells in terms of interior angles and edge length (Figure 4E-F). Therefore, the constraint not only improved the agreement of our model with the von Neumann-Mullins law, but also enabled our model to simulate the angle dynamics. Below, all the simulation results are based on the model with the angle constraint unless otherwise indicated.

According to Lewis law, the size of a cell increases with its edge number (2, 3). Based on this and our previous study (14), relaxation increases the area of large cells and decreases the area of small cells, agreeing with the predictions of the von Neumann-Mullins law. It is worth mentioning that these predictions are only correct on a statistical average, and the variation pattern of different-sized cells with the same edge number remains unknown. For this purpose, this study relaxed the random-disordered Voronoi networks with $k = 0.2$ (where all inner cells are hexagonal). The results show that small inner cells tend to expand and large inner cells tend to shrink (Figure 5A). This trend holds for different polygon classes as well as for marginal cells across the three $k$ values (Figure 5B-F). This trend is opposite to the overall trend predicted by the classical von Neumann-Mullins law. To more accurately describe the area change, this study combined the topological and geometric effects:

$$dA/dt = C_1(n - \langle n \rangle) + C_2\left(1 - A/A(n)\right), \quad (5)$$

here, $\langle n \rangle$ is the average edge number of inner or marginal cells, $A(n)$ is the preferred area of $n$-edged cells, $A$ is the cell area, and $C_1$ and $C_2$ control the topological and geometric contributions, respectively.

We compared the fitting performance of Eqs. (4) and (5) for inner cells under the three $k$ values, and found that Eq. (5) decreased the root mean square error (RMSE) by at least 19% and increased $R^2$ (coefficient of determination) by more than 26% (Figure 6). For marginal cells, RMSE and $R^2$ were also improved by more than 20% at $k = 0.6$ and by more than 11% at $k = 1$, while the improvements in RMSE and $R^2$ decreased to about 7% and 4% at $k = 0.2$, respectively (Figure 6). Therefore, for both inner and marginal cells, the addition of the geometric correction term significantly improved the fitting compared with the classical version of the von Neumann-Mullins law. In these analyses, the initial average area of $n$-edged Voronoi cells was set as $A(n)$. $C_1$ was fitted via Eq. (4), and $C_2$ was subsequently determined by minimizing the RMSE according to Eq. (5). Notably, this study reports that the ratio $C_2 : C_1$ of inner cells remains approximately constant at 2.3, independent of $k$ value and number of iterations (effects of 5 and 25 iterations on area variations were additionally tested). As for marginal cells, the ratio varied significantly around 1.8 (in the range of 1.4~2.3). In the absence of the angle constraint, Eq. (5) still provides a better fit than Eq. (4), with the average ratio $C_2 : C_1$ for inner and marginal cells being 3.7 and 3.1, respectively. Given that the geometric correction term remains effective without this constraint, we build upon this work and previous studies (11, 14, 27) to propose that Eq. (5) may be derived by introducing the central angle symmetry into the vertex motion equation.

***3.3 Aboav-Weaire law***

The Aboav-Weaire law indicates that the total edge number $mn$ of a cell's neighbouring cells increases with its edge number $n$, while Lewis law states that the area $A$ of a cell increases with $n$ (2, 3). These two laws describe different aspects of the space filling pattern on a 2D plane, yet the corresponding mathematical mechanisms are still unclear and their equations remain under debate. In our simulations, relaxation changes only the cell geometry and does not change topological features, including $n$ and $mn$. In this way, we prefer to use the following equation to analyze the neighbouring relationship of inner cells

$$mn = 5n + 6 + \mu_2, \tag{6}$$

where, $\mu_2$ is the second moment of edge numbers and calculated as $\mu_2 = \sum n^2 P_n - 36$, $P_n$ denotes the proportion of $n$-edged cells in the network. This study generated about 150,000 inner cells using software R with the package *deldir* based on the same method as described in the Method Section at $k = 1$, and found a $\mu_2$ of 0.93, which is significantly lower than the value of 1.70 reported for random Voronoi networks (14).

This study found that, for a given $n$, the average $mn$ value of inner cells is higher than that of marginal cells by about 9.4 (Figure 7A), whereas the $mn$ difference for *Pyropia* tissues is about 9.2 (6). The $mn$ difference should be mainly caused by the specific positions of marginal cells. The average edge number $\langle n \rangle$ of marginal cells is significantly lower than that of inner cells by approximately one, and marginal cells can be considered as a specific case of inner cells (6). By treating marginal cells as if they were inner cells, $\mu_2$ of inner cells can be used to predict the $mn$ of marginal cells via Eq. (6). Under this assumption, each marginal cell has its edge number increased by one and gains two additional 6-edged neighbours. We thus obtain the modified Aboav-Weaire law for an $n$-edged marginal cell

$$mn = 5(n + 1) + 6 + \mu_2 - 14 = 5n - 3 + \mu_2. \tag{7}$$

In this way, the difference between Eqs. (6) and (7) is nine, which is very close to the observed values in this and a previous study (6). Thus, the major reason for the discrepancy in $mn$ values is the difference in positions between inner and marginal cells. To evaluate the prediction performance of Eqs. (6) and (7), we adopted linear fitting combined with relative RMSE to test prediction quality (Figure 7B). The $R^2$ values of both cell types are higher than 0.88. The relative RMSE is 4.6% for inner cells and 7.0% for marginal cells, respectively. These results indicate Eqs. (6) and (7) can reliably predict the measured values with satisfactory accuracy.

### *3.4 Lewis law*

The average areas of both Voronoi and relaxed cells increase with $n$, although their slopes differ significantly; this trend holds for both inner and marginal cells (Figure 7C-D). While these results are consistent with Lewis law, distinct equations are required for Voronoi and relaxed cells to accurately describe the relationship between area and $n$. Based on the ellipse packing hypothesis (28, 29), the Voronoi cells and corresponding relaxed cells can be considered as ellipses' inscribed polygons (EIP) and EMIP, respectively. Then we examined the prediction qualities of the following equations for Voronoi cells and relaxed cells, respectively:

$$A = 0.5nabsin\left(\frac{2\pi}{n}\right)(1 - \mu_2/n), \quad (8)$$

$$A = 0.5nabsin\left(\frac{2\pi}{n}\right), \quad (9)$$

where $a$ and $b$ are the semi-long axis and semi-short axis of the fitting ellipse of an $n$-edged cell. Eq. (9) gives the ellipse's maximum inscribed polygon (EMIP) (30). The average edge number of marginal cells is less than that of inner cells by approximately one (6), and this topological difference has been incorporated in the boundary versions of equations for the von Neumann-Mullins law and Aboav-Weaire law. The average edge number $\langle n \rangle = 5$ for marginal cells appears in the boundary version of the von Neumann-Mullins law (25, 26); for the Aboav-Weaire law, this study suggests that the specific $mn$ of marginal cells originates from their specific edge numbers.

Whether $\langle n \rangle = 5$ for marginal cells needs to be considered in Eqs. (8) and (9) remains unknown. Then, this study compared the prediction quality of Eqs. (8) and (9) for marginal cells at $k = 1$ with and without the consideration of the decreased $\langle n \rangle$ by one. We found that, for Voronoi marginal cells, replacing $n$ with $n + 1$ results in an increase in relative RMSE from 31% to 41% and a decrease in $R^2$ from 0.61 to 0.56, whereas for relaxed marginal cells, the replacement leads to an increase in relative RMSE from 29% to 35% and a decrease in $R^2$ from 0.86 to 0.84. These results suggest that the boundary effect on edge number does not apply to Lewis law.

Based on linear fitting and relative RMSE (Figure 7E-F), the prediction quality of Eq. (9) is better than that of Eq. (8). For inner cells at $k = 1$, the relative RMSE values of Eqs. (8) and (9) are 20.0% and 7.4%, respectively; while the corresponding $R^2$ values are 0.89 and 0.97. Considering that the ellipse fitting method is designed for better fitting (16) rather than for finding the smallest circumscribed ellipse, the prediction quality could potentially be improved. Furthermore, when regressed against measured and predicted values, the nonlinear Eqs. (8) and (9) outperform the classical linear form of Lewis law in terms of $R^2$, regression slope and relative RMSE (data not shown). Removal of the angle constraint increased the relative RMSE of Eq. (9) from 7.4% to 14.9% and decreased the $R^2$ from 0.97 to 0.89. Thus, the angle constraint promotes the relaxation in terms of Lewis law. Besides, it is worth to mention that the prediction qualities of Eqs. (8) and (9) increase significantly with the decrease of $k$ value.

According to Eq. (8), the area of a Voronoi inner cell is $(1 - \mu_2/n)$ times that of EMIP. Then, the average area of inner and marginal cells should be about 0.84 and 0.81 times that of EMIP, respectively, based on the edge number distribution of the Voronoi network at $k = 1$. These results are consistent with our previous studies (6, 28), where the average area of *Pyropia* inner and marginal cells was about 0.82~0.9 times the EMIP.

### *3.5 Implications for geometric and topological dynamics*

Our previous study found that, for inner cells, the variation in perimeter is about twice that in area (14). This is confirmed by the present study on both inner and marginal cells, although the marginal cells with 3 and 4 edges are excluded from the fitting (Figure 8A-B). This study then quantified the effects of relaxation on shape index (SI) and found an interesting phenomenon that the SI variation is size dependent: large cells showed little variation, whereas small cells changed a lot (Figure 8C-D). Our previous study reported that relaxation increases the average edge length of large inner cells and decreases that of small inner cells (14), the present study confirmed this finding for both cell types (Figure 8E-F). Since SI depends only on shape and does not change under scaling and rotation (1), our results indicate that relaxation tends to enlarge large cells but shifts small cells toward EMIPs. This phenomenon likely stems from the fact that, compared with those of small Voronoi cells, the SI values of large Voronoi cells are not only closer to those of regular polygons, but also exhibit less variation.

Here, we discuss the possible mechanism for T1 transition in 2D foams. In 2D foams and epithelial tissues, a T1 event happens when an edge shrinks to zero, resulting in an unstable vertex with a coordination number of four (2, 8). That is why vertex models generally set the precondition for triggering of a T1 transition as an edge being shorter than a threshold length (7-13). Both this study and our previous study found that relaxation concentrated the edge lengths in range of 0.5 to 0.7, meanwhile significantly decreased the percentages of shorter ($<0.3$) edges (Figure 4C, E). In our simulations, the motion of most vertices stopped after the 100th iteration at $\gamma = 0.1$, then the whole network can be considered as fully relaxed. At this state, the edge length and cell area would not change further or change very slightly. These results suggest that the symmetric relaxation tends to inhibit the occurrence of T1.

A T1 event generally involves four cells, resulting in two cells each gaining an edge while the other two each lose an edge, e.g. T1 events occur among inner cells of 2D foams (Figure 9A-B) and *Drosophila* wing disc (21). However, at the boundary condition, a T1 event involves only three cells (Figure 9C). Previous studies also reported similar phenomena, e.g., epithelial wound healing in *Drosophila* wing disc (13) and flowing 2D foams (where most cells are attached to solid boundaries) in a channel (31). Thus, a T1 transition should be mainly determined by the shape of three neighbouring cells. Based on Lewis law and Aboav-Weaire law, small cells tend to neighbour with large cells and vice versa. The von Neumann-Mullins law predicts that large cells tend to be larger and small cells tend to be smaller. Based on these three laws, the area disparity between neighbouring cells will increase by relaxation. Consequently, to satisfy the Lewis law, smaller cells need to lose edges and larger cells need to gain edges. This analysis agrees with a recent numerical study based on the viscous froth model, which shows that, for flowing 2D foams in a channel, greater cell size disparity strongly promotes T1 events, all of which involve only three cells (32).

The von Neumann-Mullins law has been observed in 2D foams and simulations via vertex model in the presence of T1 and T2 (10, 11, 33). The coexistence of the von Neumann-Mullins law and T1 indicates that understanding the relationship between the effects of symmetric relaxation on cell area and edge length is the key to explaining the mechanism of T1 in 2D foams. Based on the above analyses, this study proposes that the induction of T1 can be divided into two phases: first, symmetric relaxation shifts the cell shape toward EMIP, manifesting as an increase in area disparity between neighbouring cells and a decrease in the proportion of shorter edges; second, force disequilibrium increases with area disparity and finally overcomes the effects of symmetric relaxation on edge length, triggering the shrinkage of an edge of the smaller cell. Based on the present study (Figure 8E-F) and our previous study (14), relaxation increasing the average edge length of large cells and decreasing that of small cells. Therefore, the shrinkage target should not be the edge shared by a large cell and a small cell, but the other edge of the small cell, e.g., the edge shared by two smaller cells (Figure 9D). Given that some consequences of topological transitions in 2D foams and living 2D networks are opposite, e.g. effects of different topological transitions (T2 vs. cell division) on cell numbers, size-dependent edge gain/loss in T1 events (14, 21, 33), the T1 transition in living 2D networks may proceed in an opposite manner compared with that in 2D foams.

## 4 Conclusion

This study developed a symmetric relaxation method for entire 2D cellular networks. Specifically, the vertex movement is governed by the central angle symmetry of associated cells and the angle symmetry at each vertex. Inner vertices are main controlled by the former, while marginal vertices by the latter. Moreover, because the central angle symmetry couples all vertices of each cell, it imposes a strong constraint on cell shape variation. The relaxation simulation is applied to trimmed Voronoi networks with different irregularities, which are used as the initial networks. The main findings are as follows:

We find that the simulation results of both inner and marginal cells agree with von Neumann-Mullins law, and a modified equation of the law that includes a geometric correction term (Eq. 5) improves the prediction quality, indicating that both topological and geometric factors contribute to area changes. The Aboav-Weaire law for marginal cells can be derived by assuming them as inner cells, leading to Eq. 7. We also find that relaxation shifts cells toward ellipses' maximum inscribed polygons (EMIP), which agrees with the ellipse packing hypothesis. In this study, symmetric relaxation reduces the proportion of very short edges and increases the area disparity between neighbouring cells. We then propose a two-phase induction for the T1 transition: relaxation first drives cells toward EMIP, increasing area disparity but decreasing the proportion of short edges; then, when the force disequilibrium becomes large enough, an edge shared by two smaller cells shrinks and triggers a T1 transition.

Overall, this study provides a symmetry-based tool for equilibrating entire 2D cellular networks and reproduces key empirical laws, thereby offering insights into the induction mechanisms of T1 transitions from the perspective of symmetric relaxation.

## Funding

No funding was received.

## Declaration of interest statement

No potential conflict of interest was reported by the authors.

**Figure 1** Diagrams of symmetric relaxation for inner and marginal vertices. (A) Relaxation moves each inner vertex $V$ toward its target position $P$, which is the centroid of a triangle formed by three optimal rays corresponding to the vertex (14). The movement is also restricted by the angle symmetry at each vertex. (B) For each marginal vertex $V$, relaxation moves it along the marginal edge associated with its smaller marginal angle, and the midpoint of this edge is the target position $P$. The movement is also influenced by the central angle symmetry of two associated marginal cells. Green and purple dashed lines represent the optimal rays and interior angles, respectively. $D$ is the distance between the initial and target positions of the vertex.

**Figure 2** Trimmed Voronoi networks and their corresponding relaxed networks with and without the angle constraint. The Voronoi networks are generated at three irregularity ($k$) values.

**Figure 3** Edge number ($n$) distributions of inner (A) and marginal (B) cells of typical 2D cellular networks: amorphous $SiO_2$ network (17), *Pyropia* thallus (6, 18), larval wing disc of *Drosophila* (19-21), and trimmed Voronoi networks. IC: inner cells; MC: marginal cells.

**Figure 4** Effect of relaxation on cell area, edge length, and interior angles. Relationships between area change $\Delta A$ (difference between the Voronoi cells and their fully relaxed counterparts) and edge number $n$ of inner (A) and marginal (B) cells at three $k$ values with angle constraint and at $k = 1$ without angle constraint. Distributions of edge length (C) and interior angles (D) of inner cells at $k = 1$. Distributions of edge length (E) and interior angles (F) of marginal cells at $k = 1$. IC: inner cells; MC: marginal cells.

**Figure 5** Relationships between the initial area $A$ and area change $\Delta A$ for inner and marginal cells across three $k$ values. IC: inner cells; MC: marginal cells.

**Figure 6** Linear fitting between observed $\Delta A$ and predicted $\Delta A$ for inner and marginal cells based on Eqs. (4) and (5) across three $k$ values. IC: inner cells; MC: marginal cells.

**Figure 7** Observed data of $mn$ and area $A$ for inner and marginal cells, and the corresponding linear fittings between observed and predicted data. IC: inner cells; MC: marginal cells. Data for IC and MC are combined due to their highly identical linear fitting slopes.

**Figure 8** Relationships between geometric variations of inner and marginal cells. (A, B) Perimeter change $\Delta P$ versus area change $\Delta A$. (C, D) Shape index change $\Delta SI$ versus initial area $A$. (E, F) Change in cell's average edge length $\Delta E_{avg}$ versus edge number $n$. Three- and four- edged marginal cells are excluded from the fitting in (B). IC: inner cells; MC: marginal cells.

**Figure 9** Observed T1 events in 2D foams (A-C) and a schematic diagram for T1 in 2D foams (D). Cell $a$ represents the largest among the cells involved in the T1 event. Cell $d$ can be smaller (A) or larger (B) than cells $b$ and $c$, whereas under boundary conditions, only three cells participate in the event (C). A T1 event results in the large cell $a$

gaining an edge, while two small cells $b$ and $c$ each lose an edge (D). The red line represents the edge shared by two small cells, which shrinks to zero during the T1 event.

Figure 1

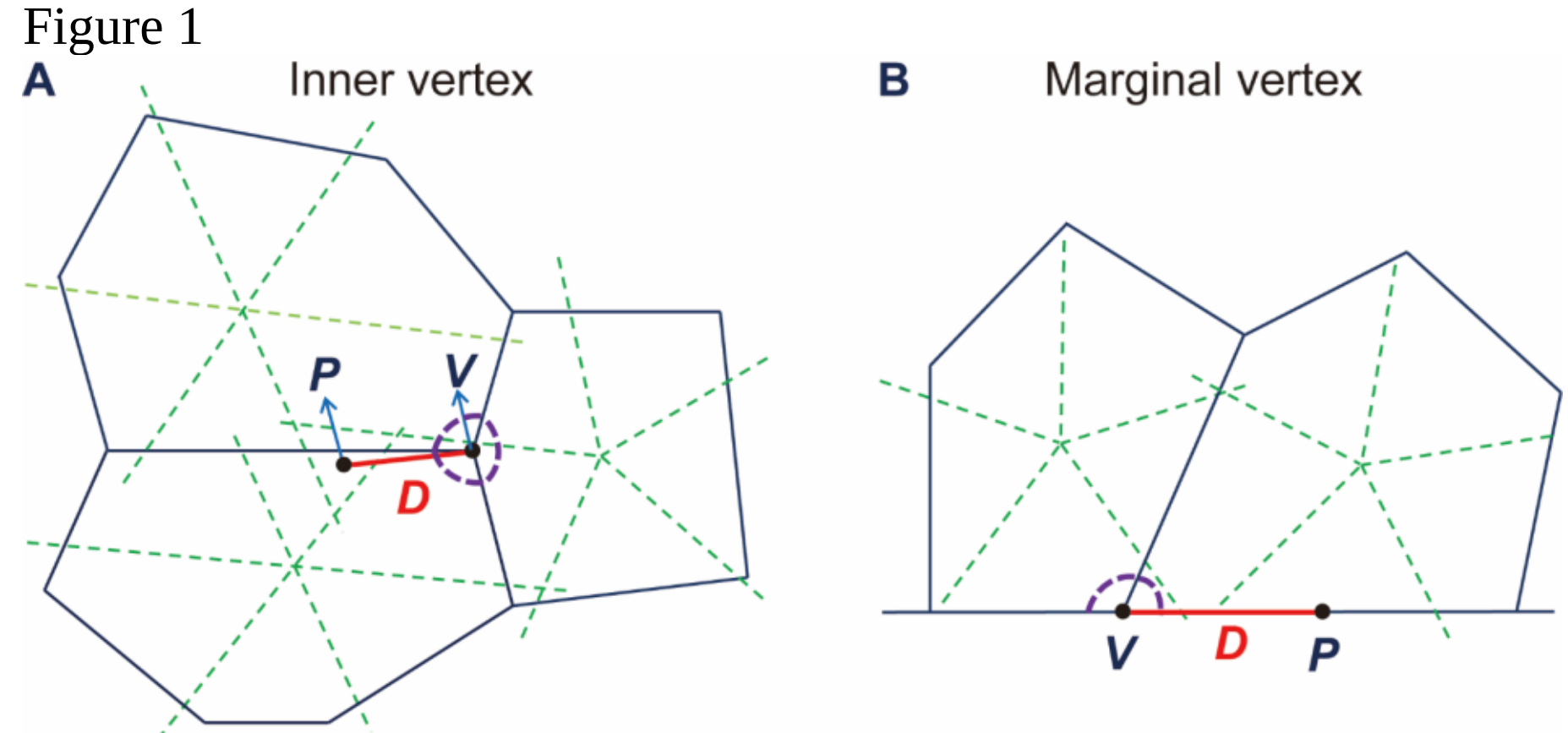

Figure 2

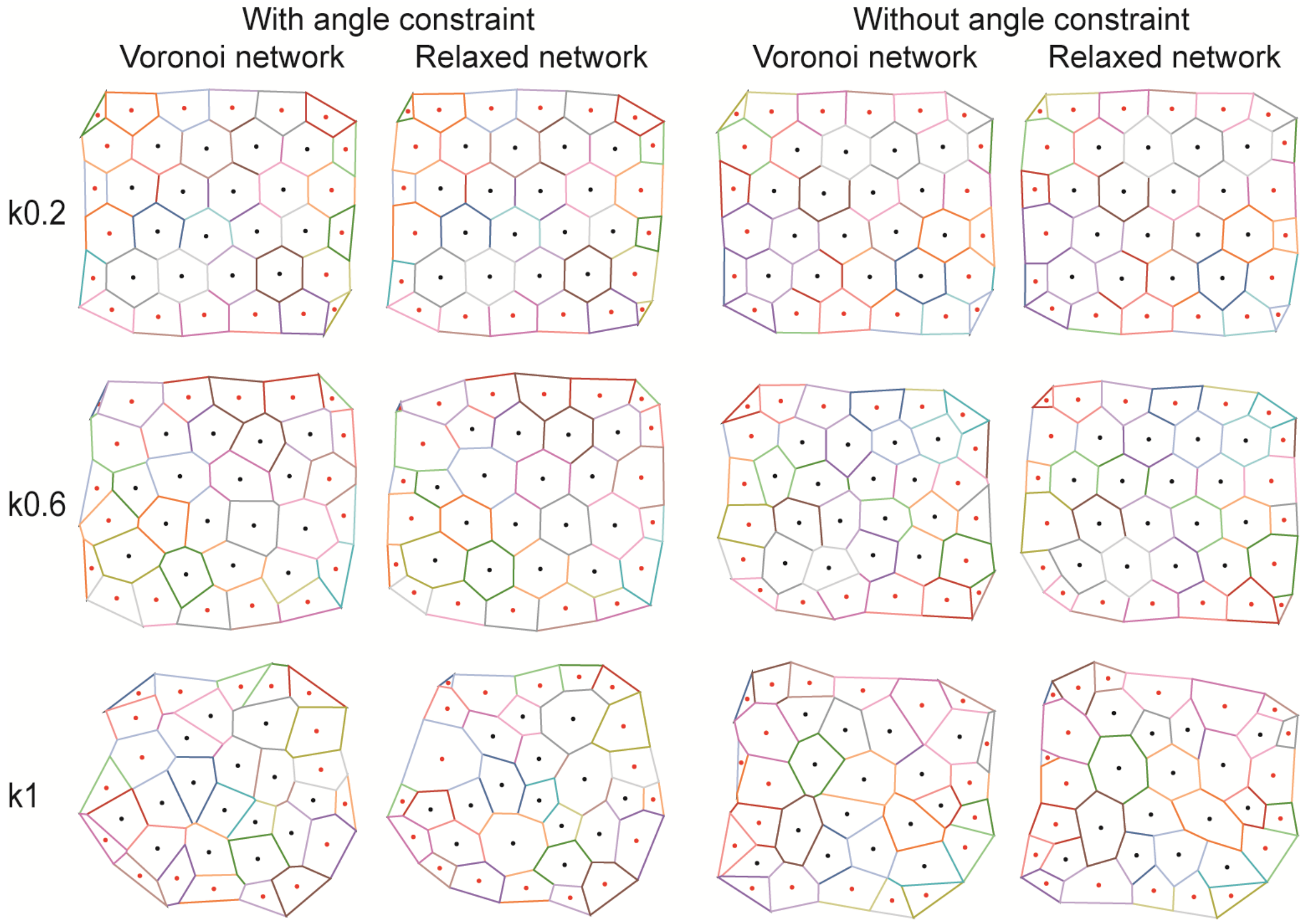

With angle constraint
Voronoi network
Relaxed network
Without angle constraint
Voronoi network
Relaxed network
k0.2
k0.6
k1

Figure 3

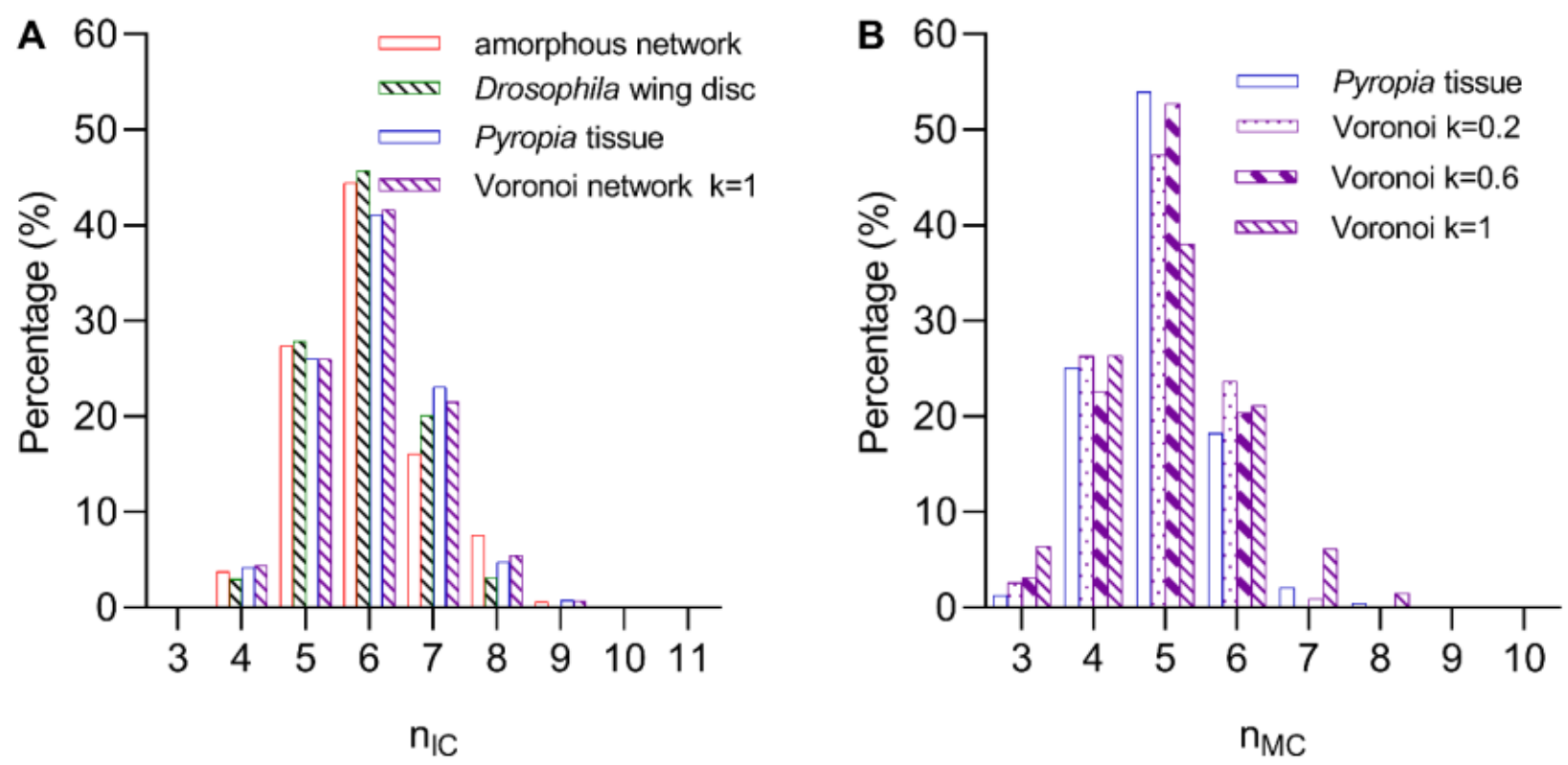

A
B
Percentage (%)
amorphous network
Drosophila wing disc
Pyropia tissue
Voronoi network k=1
Voronoi k=0.2
Voronoi k=0.6
Voronoi k=1
$n_{IC}$
$n_{MC}$

Figure 4

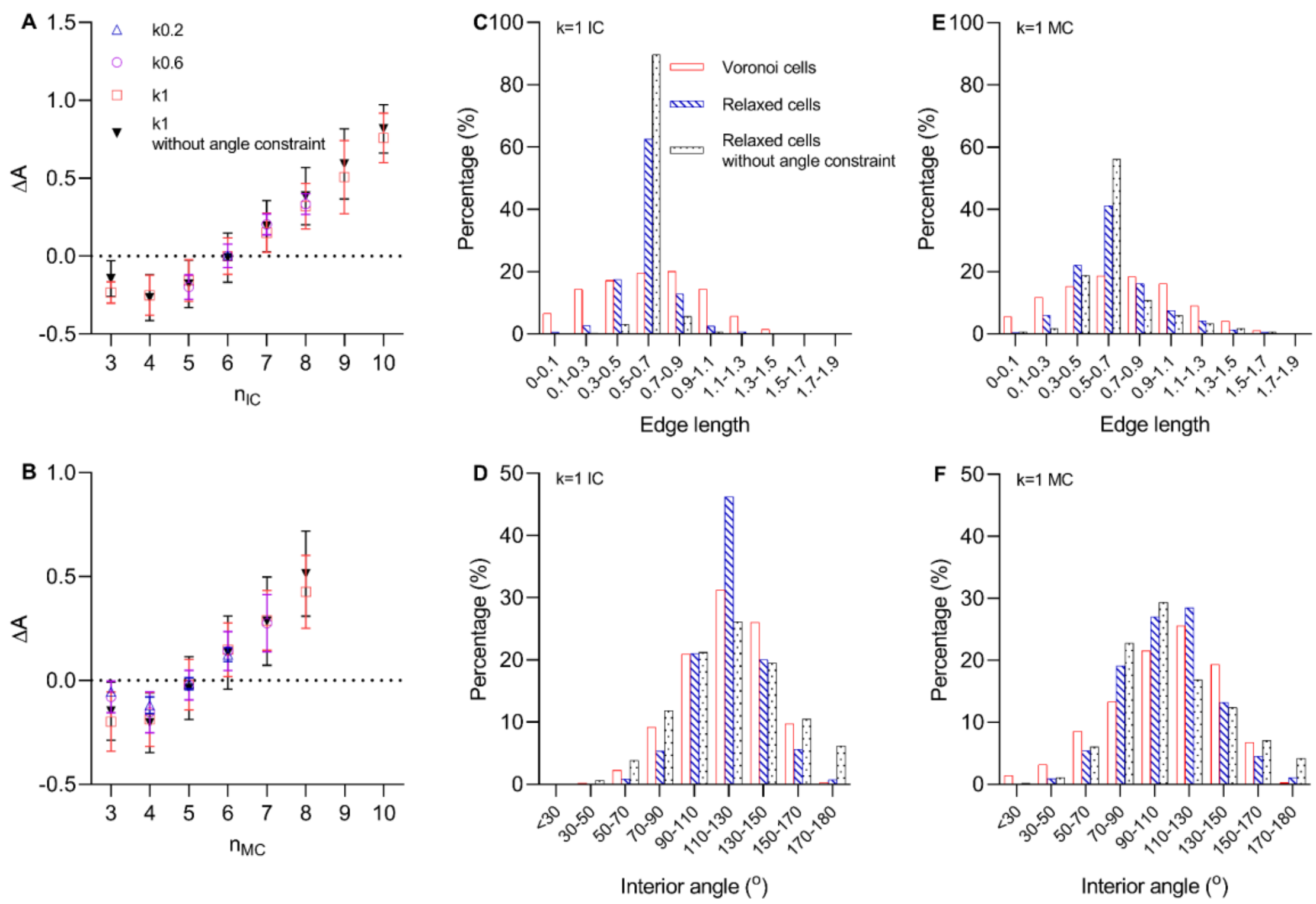

A
k0.2
k0.6
k1
k1
without angle constraint
ΔA
$n_{IC}$
B
$n_{MC}$
C
k=1 IC
Voronoi cells
Relaxed cells
Relaxed cells
without angle constraint
Percentage (%)
Edge length
E
k=1 MC
D
Interior angle (°)
F

Figure 5

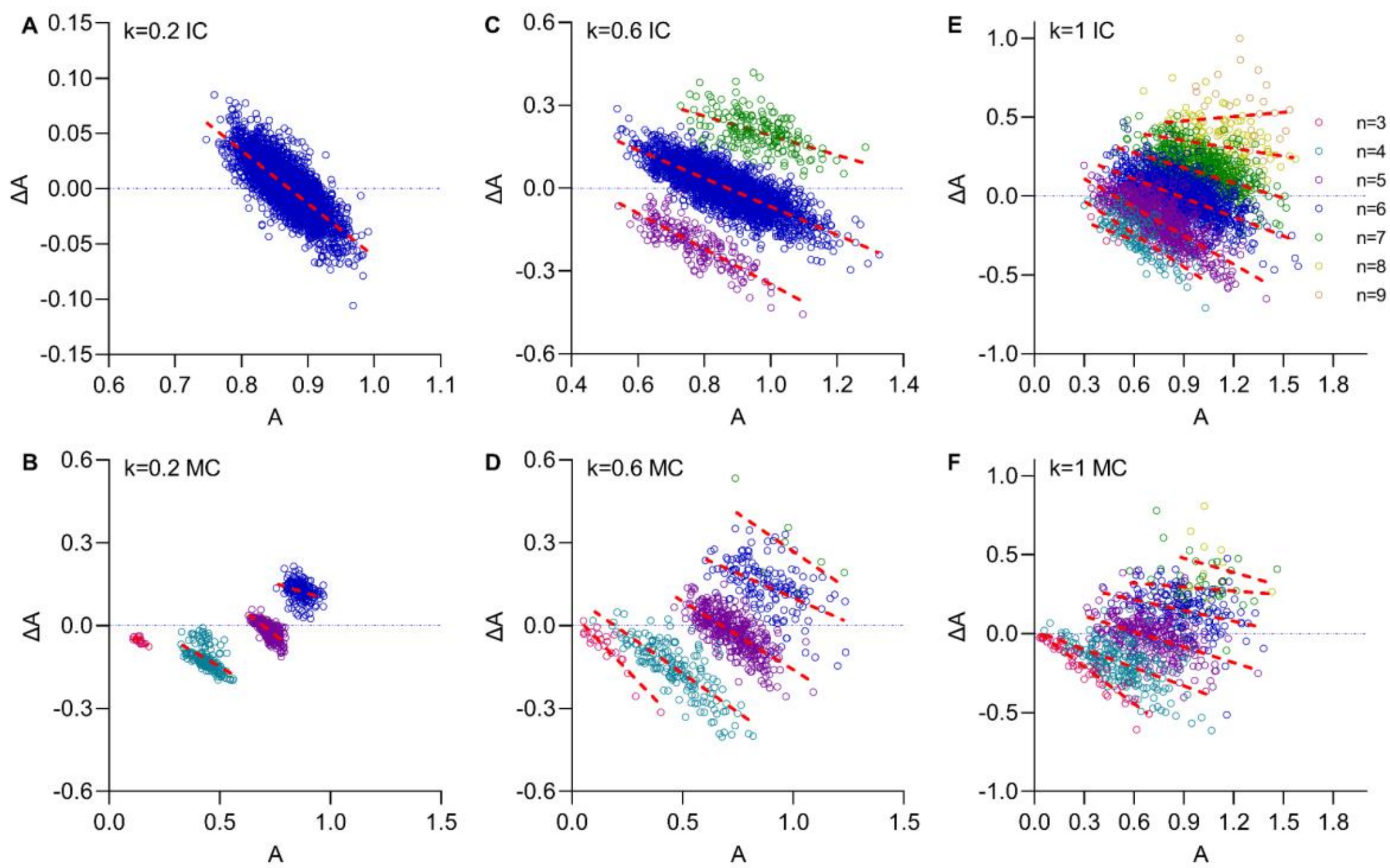

A
k=0.2 IC
B
k=0.2 MC
C
k=0.6 IC
D
k=0.6 MC
E
k=1 IC
F
k=1 MC
ΔA
A
n=3
n=4
n=5
n=6
n=7
n=8
n=9

Figure 6

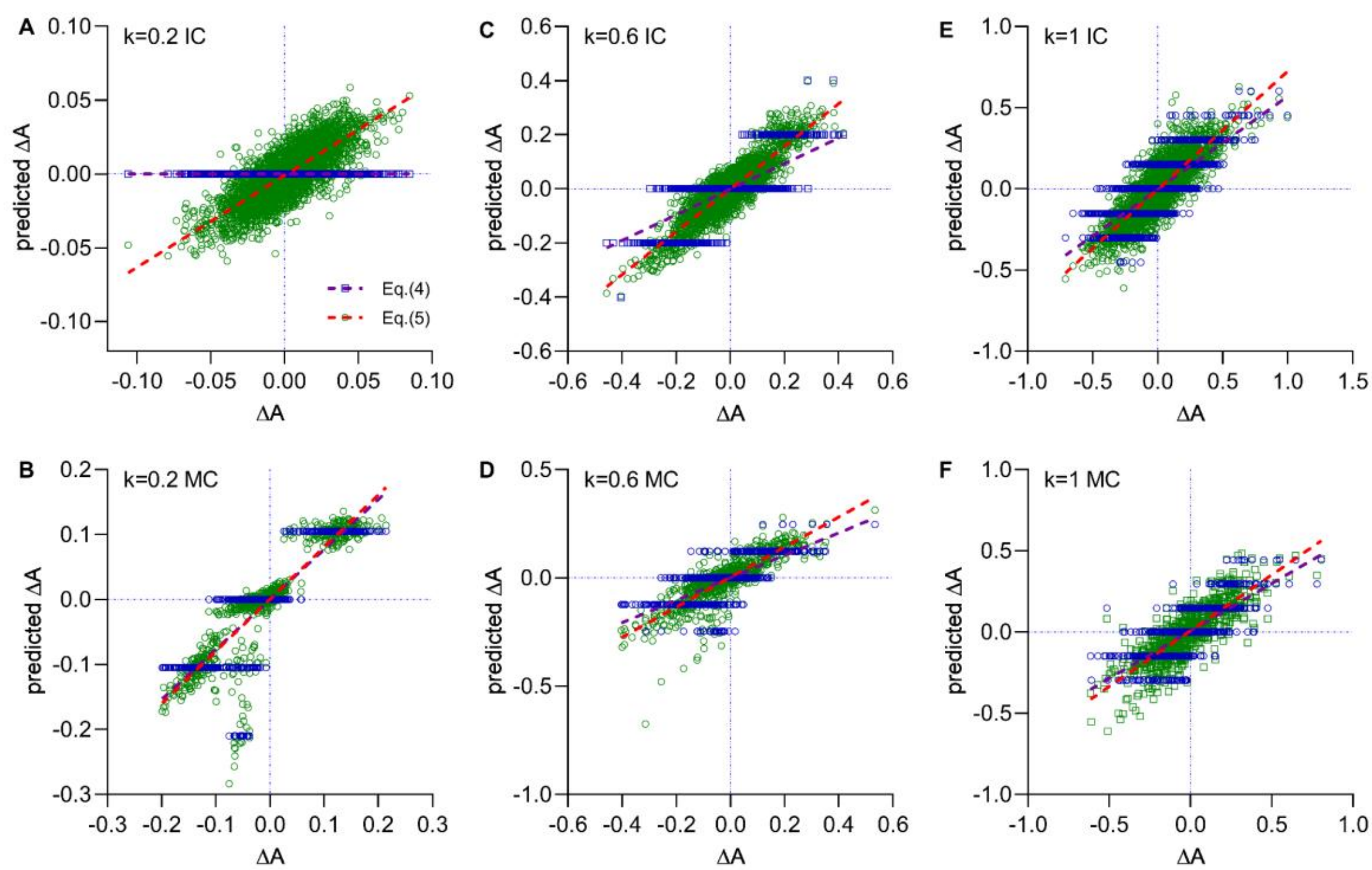

A
k=0.2 IC
predicted ΔA
ΔA
Eq.(4)
Eq.(5)
B
k=0.2 MC
C
k=0.6 IC
D
k=0.6 MC
E
k=1 IC
F
k=1 MC

Figure 7

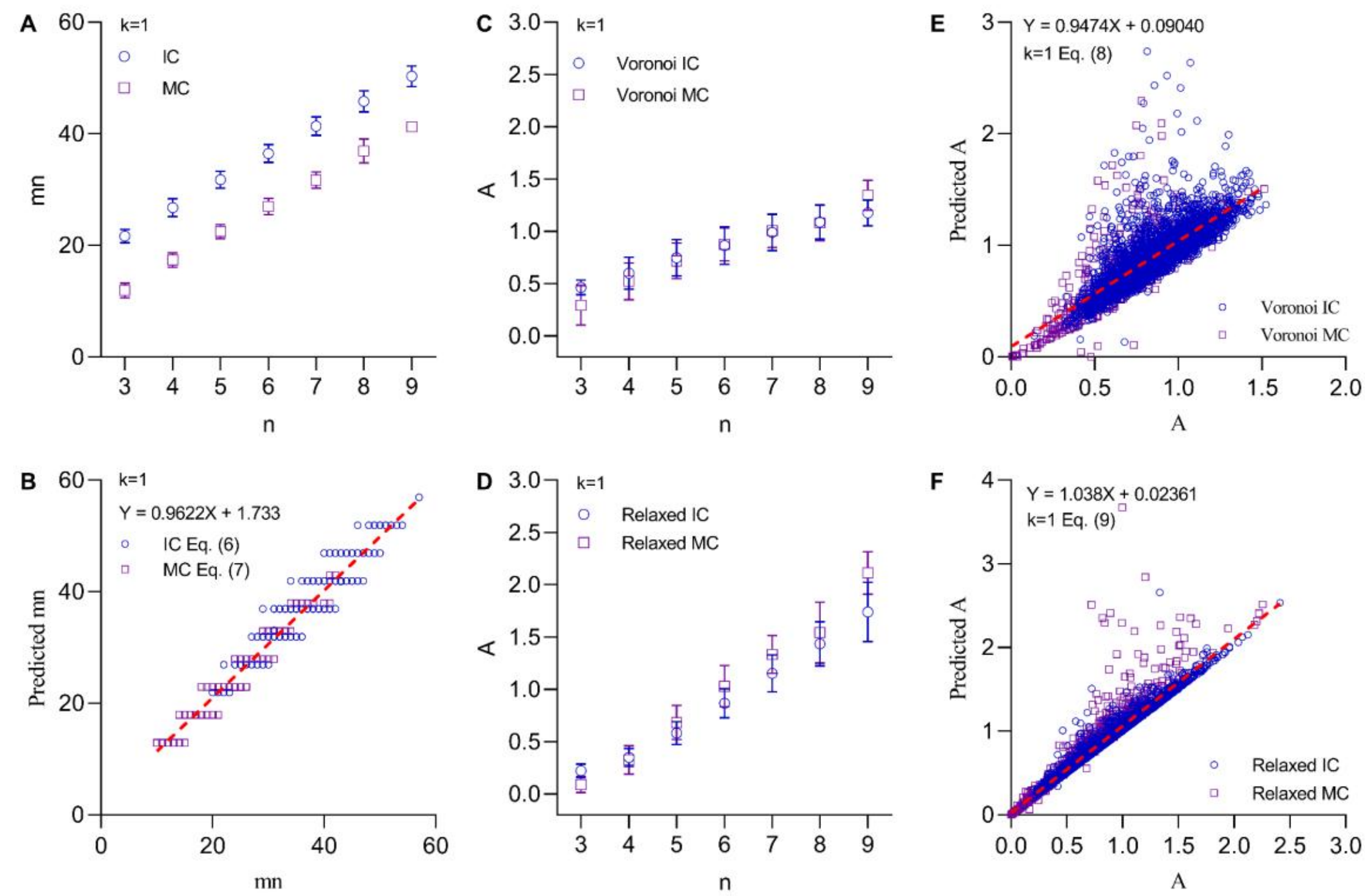
A
k=1
IC
MC
mn
n
B
k=1
Y = 0.9622X + 1.733
IC Eq. (6)
MC Eq. (7)
Predicted mn
mn
C
k=1
Voronoi IC
Voronoi MC
A
n
D
k=1
Relaxed IC
Relaxed MC
A
n
E
Y = 0.9474X + 0.09040
k=1 Eq. (8)
Voronoi IC
Voronoi MC
Predicted A
A
F
Y = 1.038X + 0.02361
k=1 Eq. (9)
Relaxed IC
Relaxed MC
Predicted A
A

Figure 8

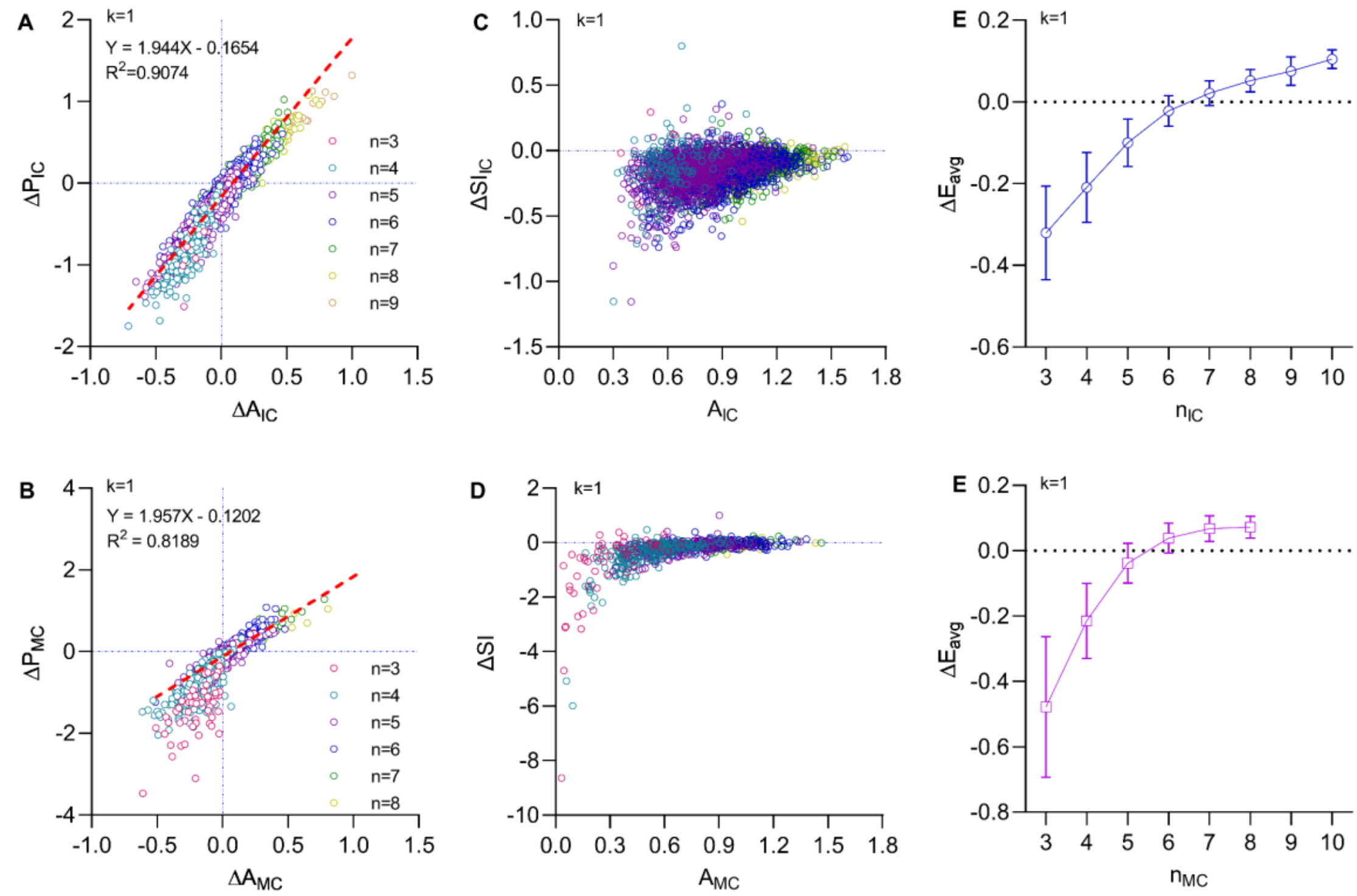
A
k=1
Y = 1.944X - 0.1654
R2=0.9074
n=3
n=4
n=5
n=6
n=7
n=8
n=9
ΔP_IC
ΔA_IC
C
k=1
ΔSI_IC
A_IC
E
k=1
ΔE_avg
n_IC
B
k=1
Y = 1.957X - 0.1202
R2 = 0.8189
n=3
n=4
n=5
n=6
n=7
n=8
ΔP_MC
ΔA_MC
D
k=1
ΔSI
A_MC
E
k=1
ΔE_avg
n_MC

Figure 9

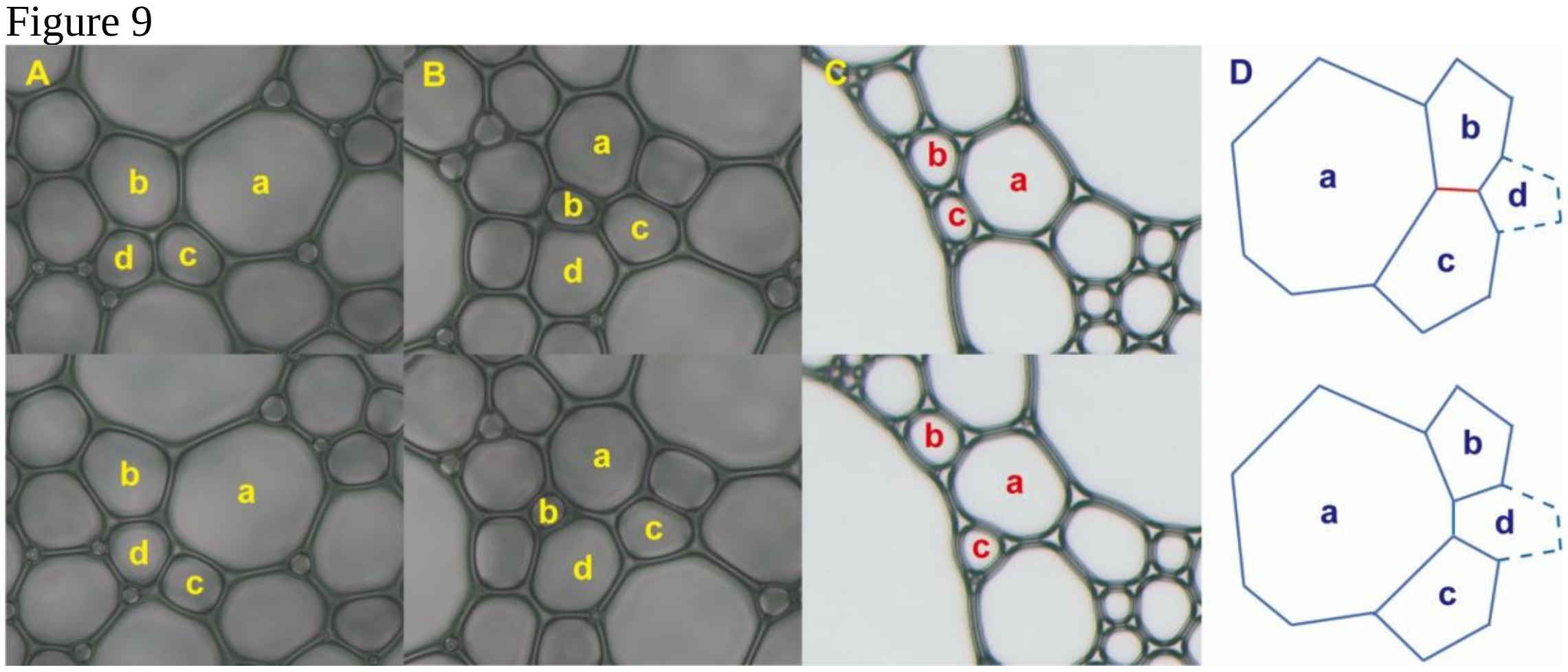